\documentclass[prx,letterpaper,aps,superscriptaddress,floatfix,twocolumn,nofootinbib]{revtex4-2}
\pdfoutput=1
\usepackage{graphicx}
\usepackage{amsmath}
\usepackage{amsfonts}
\usepackage{amssymb}
\usepackage[normalem]{ulem}
\usepackage[small,bf]{subfigure}
\usepackage[utf8]{inputenc}
\usepackage{dsfont}
\usepackage{xcolor}
\usepackage{slashed}
\usepackage{soul}
\usepackage{comment}
\usepackage{tikz-cd}
\usepackage{hyperref}

\newcommand{\beq}{\begin{equation}}
\newcommand{\eeq}{\end{equation}}

\newcommand{\beqa}{\begin{eqnarray}}
\newcommand{\eeqa}{\end{eqnarray}}

\usepackage{amsthm}
\theoremstyle{plain}
\newtheorem*{theorem*}{Theorem}
\newtheorem{theorem}{Theorem}
\newtheorem{corollary}{Corollary}[theorem]
\newtheorem{manualtheoreminner}{Theorem}
\newenvironment{manualtheorem}[1]{
  \renewcommand\themanualtheoreminner{#1}
  \manualtheoreminner
}{\endmanualtheoreminner}

\begin{document}
\title{Non-zero momentum requires long-range entanglement
}
\author{Lei Gioia}
\affiliation{Perimeter Institute for Theoretical Physics, Waterloo, Ontario N2L 2Y5, Canada}
\affiliation{Department of Physics and Astronomy, University of Waterloo, Waterloo, Ontario 
N2L 3G1, Canada} 
\author{Chong Wang}
\affiliation{Perimeter Institute for Theoretical Physics, Waterloo, Ontario N2L 2Y5, Canada}
\begin{abstract}
We show that a quantum state in a lattice spin (boson) system must be long-range entangled if it has non-zero lattice momentum, i.e. if it is an eigenstate of the translation symmetry with eigenvalue $e^{iP}\neq1$. Equivalently, any state that can be connected with a non-zero momentum state through a finite-depth local unitary transformation must also be long-range entangled. The statement can also be generalized to fermion systems. Some non-trivial consequences follow immediately from our theorem: (1) several different types of Lieb-Schultz-Mattis-Oshikawa-Hastings (LSMOH) theorems, including a previously unknown version involving only a discrete $\mathbb{Z}_n$ symmetry, can be derived in a simple manner from our result; (2) a gapped topological order (in space dimension $d>1$) must \textit{weakly} break translation symmetry if one of its ground states on torus has nontrivial momentum -- this generalizes the familiar physics of Tao-Thouless; (3) our result provides further evidence of the ``smoothness'' assumption widely used in the classification of crystalline symmetry-protected topological (cSPT) phases.
\end{abstract}
\maketitle

\section{Introduction}
\label{sec:intro}

The ubiquitous appearance of translation symmetry in physical systems signals the importance of having a complete picture of the complex role it may play. In particular, although the ground state energy (associated with time-translation symmetry) of a many-body quantum system or a quantum field theory is frequently studied, the ground state \textit{momentum} (associated with space-translation symmetry) is rarely discussed. Rather, in most cases one focuses on the momentum difference between excited states and the ground state. In this work we reveal a connection between the momentum and the entanglement structure of a quantum state, in the context of lattice spin (boson) systems:
\begin{theorem}
If a quantum state $|\Psi\rangle$ in a lattice spin (boson) system is an eigenstate of the lattice translation operator $T:|\Psi\rangle\to e^{iP}|\Psi\rangle$ with a non-trivial momentum $e^{iP}\neq1$, then $|\Psi\rangle$ must be long-range entangled, namely $|\Psi\rangle$ cannot be transformed to an un-entangled product state $|000...\rangle$ through an adiabatic evolution or a finite-depth quantum circuit (local unitary).

\label{Thm}
\end{theorem}

The intuition behind this statement follows from the sharp difference between translation $T$ and an ordinary onsite symmetry $G$ that is defined as a tensor product of operators acting on each lattice-site (such as the electromagnetic $U(1)$). A product state may recreate any total symmetry charge $Q$ under $G$ by simply assigning individual local Hilbert space states to carry some charge $Q_\alpha$ such that $Q=\sum_{\alpha}Q_\alpha$. However in the case of non-onsite translation symmetry, all translation-symmetric product states, which take the form $|\alpha\rangle^{\otimes L}$, can only carry trivial charge (lattice momentum). This suggests that non-trivial momentum is an inherently non-local quantity that cannot be reproduced without faraway regions still retaining some entanglement knowledge of each other, i.e. the state must be long-range entangled.

In condensed matter physics, we are often interested in ground states of translational-invariant local Hamiltonians. If the ground state is short-range entangled~\cite{PhysRevB.82.155138} (SRE) in the sense that it is connected to a product state through a finite-depth (FD) quantum circuit, then we expect the ground state to be unique, with a finite gap separating it from the excited states. In contrast for long-range entangled~\cite{PhysRevB.82.155138} (LRE) ground states, we expect certain ``exotic'' features: possible options include spontaneous symmetry-breaking cat states (e.g. GHZ-like states), topological orders (e.g. fractional quantum Hall states), and gapless states (e.g. metallic or quantum critical states). Theorem~\ref{Thm} provides us an opportunity to explore the interplay between translation symmetry and the above modern notions. An immediate corollary is
\begin{corollary}
If a non-zero momentum state $|\Psi\rangle$ is realized as a ground state of a local spin Hamiltonian, then the ground state cannot be simultaneously unique and gapped. Possible options include (1) gapless spectrum, (2) intrinsic topological order and (3)  spontaneous translation symmetry breaking.
\end{corollary}
In fact, we show in Sec.~\ref{sec:wCDW} that option (2) is a special subset of option (3) through the mechanism of ``weak symmetry-breaking"~\cite{Kitaev06}.

Our result is reminiscent of the celebrated Lieb-Schultz-Mattis-Oshikawa-Hastings (LSMOH) theorems \cite{LSM,Oshikawa1999,Hastings04}, which state that in systems with charge $U(1)$ and translation symmetries, a ground state with fractional $U(1)$ charge filling (per unit cell) cannot be SRE. In our case the non-trivial lattice momentum $e^{iP}\neq1$ plays a very similar role as the fractional charge density in LSMOH. In fact, as we discuss in Sec.~\ref{sec: LSMOH}, our theorem can be viewed as a more basic version of LSMOH that only involves translation symmetry, from which the standard LSMOH can be easily derived. As a by-product, we also discover a previously unknown version of LSMOH constraint that involves an onsite $\mathbb{Z}_n$ symmetry and lattice translations. 

The rest of this paper will be structured as follows: in Sec.~\ref{sec:proof} we provide a proof of Theorem~\ref{Thm} via a quantum circuit approach, and generalize it to fermion systems. Three consequences of Theorem~\ref{Thm} are discussed in Sec.~\ref{sec:consequences}: in Sec.~\ref{sec: LSMOH} we discuss several LSMOH-type theorems; in Sec.~\ref{sec:wCDW} we show that a gapped topological order must \textit{weakly} break translation symmetry if one of its ground states on torus has nonzero momentum -- this is a generalization of the Tao-Thouless physics in fractional quantum Hall effect~\cite{PhysRevB.28.1142,PhysRevB.77.155308}; in Sec.~\ref{sec:cSPT} we discuss the implication of Theorem~\ref{Thm} for the classification of crystalline symmetry-protected topological (SPT) phases. We end with some discussions in Sec.~\ref{sec:discussions}.

\section{Proof}
\label{sec:proof}

In this section we prove that SRE states necessarily possess trivial momentum, conversely implying that all non-trivial momentum ground states must be LRE. The approach that we take utilizes the quantum circuit formalism, which is equivalent to the usual adiabatic Hamiltonian evolution formulation~\cite{doi:10.1126/science.273.5278.1073,PhysRevB.82.155138} but conceptually cleaner. In particular we will harness the causal structure of quantum circuits, which will allow us to `cut and paste' existing circuits to create useful new ones.

We shall first prove Theorem~\ref{Thm} in one space dimension, from which the higher-dimensional version follows immediately.

\subsection{Proof in $1$d}
\label{sec:1dproof}

First let us specify our setup more carefully. We consider a spin (boson) system with a local tensor product Hilbert space $\mathcal{H}=\otimes_i\mathcal{H}_i$ where $\mathcal{H}_i$ is the local Hilbert space at unit cell $i$. The system is put on a periodic ring with $L$ unit cells so $i\in\{1,2...L\}$. In each unit cell the Hilbert space $\mathcal{H}_i$ is $q$-dimensional ($q$ does not depend on $i$), with a basis labeled by $\{|a_i\rangle_i\}$ ($a_i\in\{0,1...q-1\}$). The translation symmetry is implemented by a unitary operator that is uniquely defined through its action on the tensor product basis 
\beqa
\label{eq:Tboson}
    T:\hspace{5pt} & & |a_1\rangle_1\otimes|a_2\rangle_2\otimes...|a_{L-1}\rangle_{L-1}\otimes|a_L\rangle_L \nonumber \\ & &\longrightarrow |a_L\rangle_1\otimes|a_1\rangle_2\otimes...|a_{L-2}\rangle_{L-1}\otimes|a_{L-1}\rangle_L.
\eeqa
Under this definition of translation symmetry (which is the usual definition), we have\footnote{Importantly, we are not dealing with translation under twisted boundary condition, in which case $T^L=g$ for some global symmetry $g$. Many of our conclusions in this work need to be rephrased or reexamined for such twisted translations.} $T^L=1$ and any translational-symmetric product state $|\varphi\rangle^{\otimes L}$ has trivial lattice momentum $e^{iP}=1$.

Now consider a SRE state $|\Psi_{P(L)}\rangle$ with momentum $P(L)$. By SRE we mean that there is a quantum circuit $U$ with depth $\xi\ll L$ that sends $|\Psi_{P(L)}\rangle$ to the product state $|\mathbf{0}\rangle\equiv |0\rangle^{\otimes L}$ (we do not assume $U$ to commute with translation). The depth $\xi$ will be roughly the correlation length of $|\Psi_{P(L)}\rangle$. Our task is to prove that $P(L)=0$ mod $2\pi$ as long as $\xi\ll L$. Notice that this statement is in fact stronger than that for FD circuit which requires $\xi\sim O(1)$ as $L\to \infty$. For example, our result holds even if $\xi\sim {\rm PolyLog}(L)$, which is relevant if we want the quantum circuit to simulate an adiabatic evolution more accurately~\cite{Haah2018}. Our result is also applicable if the existence of $U$ requires extra ancilla degrees of freedom (DOF) that enlarges the onsite Hilbert space to $\tilde{\mathcal{H}}_i$ with dimension $\tilde{q}>q$ (for example see Ref.~\cite{ElsePoWatanabe2019}), since ancilla DOFs by definition come in product states and therefore cannot change the momentum.

The proof will be split into two steps where in \textit{Step 1} we first prove that the momentum is trivial for all $L=mn$ where $m,n\in\mathbb{Z}^+$ are mutually coprime satisfying $m,n\gg\xi$. In \textit{Step 2} we use the results of \textit{Step 1} to show that this may be extended to all other lengths.

\underline{\textit{Step 1:}} A key ingredient of the proof is to recognize that the entanglement structure of the SRE state $|\Psi_{P(L)}\rangle$ on system size $L=mn$, where $m,n\in\mathbb{Z}^+$ and $n\gg\xi$, is adiabatically connected to that of $m$ identical unentangled length $n$ SRE systems. The existence of such an adiabatic deformation, which is of a similar flavour to those presented in Refs.~\cite{PhysRevX.7.011020} and \cite{PhysRevB.96.205106}, is due to the finite correlation length of SRE systems, and will be explained in the following paragraph.

\begin{figure}[t]
    \centering
        \includegraphics[width=0.98\columnwidth]{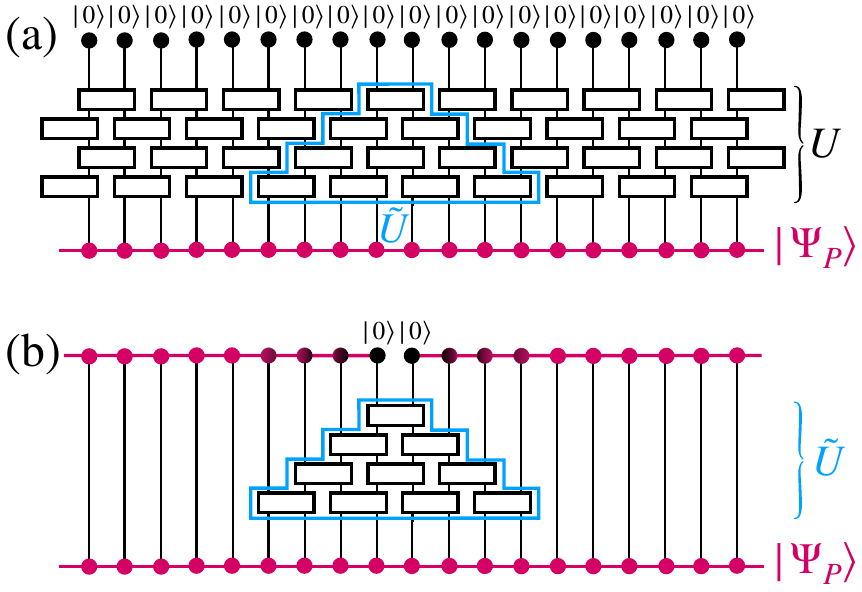}
    \caption{\label{fig:unitaries} (Color online) Depiction of finite-depth quantum circuits applied on $|\Psi_P\rangle$. Here qudits are depicted as solid circles while unitaries are depicted as rectangles. (a) A SRE state $|\Psi_P\rangle$ is always connected to the $|\mathbf{0}\rangle$ trivial state via a FD quantum circuit $U$. From $U$ a lightcone-like `adiabatic cut' $\tilde{U}$ can be created (framed in blue). (b) $\tilde{U}$ connects $|\Psi_P\rangle$ to a state that is completely decoupled across the cut.}
\end{figure}

Take the SRE state $|\Psi_{P(L)}\rangle$ placed on a periodic chain of length $L=mn$ with $m,n\in\mathbb{Z}^+$ and $n\gg\xi$. Let us try to decouple this system at some point (say between site $i$ and $i+1$) via an adiabatic evolution, creating an `open' chain.
To show that such a decoupling cut exists, we use the fact that SRE states always have a FD quantum circuit $U$ that sends the ground state to the $|\mathbf{0}\rangle\equiv |0\rangle^{\otimes L}$ product state (see Fig.~\ref{fig:unitaries}(a)). The appropriate cut is then created by modifying this circuit to form a new lightcone-like FD quantum circuit $\tilde{U}$ with all unitaries outside the `lightcone', i.e. those that do not affect the transformation that sends the two sites $i$ and $i+1$ to $|0\rangle$, set to identity (see Fig.~\ref{fig:unitaries}(b)). Such a modified circuit would span $\sim\xi$ qudits on either side of the cut and by construction takes the two sites on either side of the cut to $|0\rangle$, thus completely removing any entanglements across the link\footnote{This can be better understood in reverse: consider the state constructed by $\tilde{U} U^\dag|\mathbf{0}\rangle$ ($=\tilde{U}|\Psi_{P(L)}\rangle$) which never directly couples qudits on either side of the cut. Thus $\tilde{U}$ can be understood as completely removing the entanglement across the applied link.}.
Let us concretely take $\tilde{U}^{[0]}$ to denote the appropriate lightcone cut between the last and first qudit (recall that we are on a ring), and define the shifted adiabatic cut between the $x-1$ and $x$th qudits to be $\tilde{U}^{[x]}\equiv T^x \tilde{U}^{[0]}T^{-x}$. If the ground state is translation-symmetric we have $\tilde{U}^{[x]}|\Psi_{P(L)}\rangle=e^{-i x P(L)}T^x\tilde{U}^{[0]}|\Psi_{P(L)}\rangle$
so we see that $\tilde{U}^{[x]}$ performs the same cut (up to a phase factor) at any link. By construction this means that the local density matrices of a region surrounding the cut obeys $\rho_{lr}=\rho_l\otimes|00\rangle\langle00|\otimes\rho_r$, where the left (right) region to the cut is denoted $l$ ($r$), which in turn implies that the operations $\tilde{U}^{[x]}$ disentangles the system along that cut.

\begin{figure}[t]
    \centering
        \includegraphics[width=0.95\columnwidth]{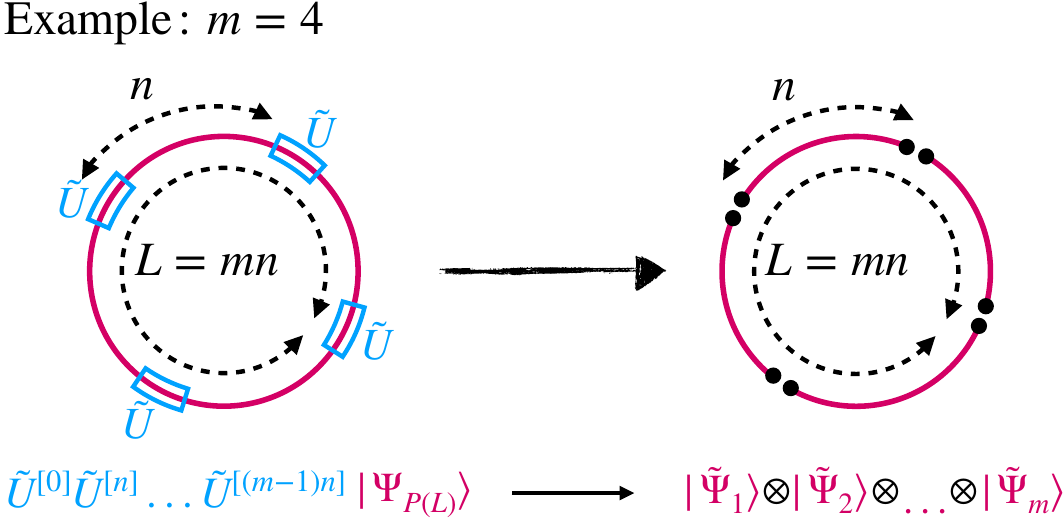}
    \caption{\label{fig:adiabaticcutting} (Color online) Illustration of the adiabatic cutting procedure on a periodic length $L=mn$ chain. Here we take $m=4$ example to demonstrate how four identical cuts, applied by $\tilde{U}$ (blue rectangle) at every $n$th link, on a length $L=4n$ state $|\Psi_{P(L)}\rangle$ (purple circle) produces four decoupled length $n$ SRE states.}
\end{figure}

The cutting procedure may be simultaneously applied to two separate links, as long as they are separated by a distance much greater than the correlation length. With this in mind, let us identically apply the cut on an $L=mn$ length system with a cut after every $n$th qudit, as depicted in Fig.~\ref{fig:adiabaticcutting}, via the FD quantum circuit $\tilde{U}^{[0]}\tilde{U}^{[n]}...\tilde{U}^{[(m-1)n]}$. Since the adiabatic deformation fully disentangles the system across the cuts, the resulting state should take the form $|\tilde{\Psi}_1\rangle\otimes|\tilde{\Psi}_2\rangle\otimes...|\tilde{\Psi}_m\rangle$ where each $|\tilde{\Psi}_i\rangle$ is an $n$-block SRE state. 

Now let us examine the symmetries of this resultant system. The original $\mathbb{Z}_{mn}$ translation symmetry, generated by operator $T$, of the original system is broken by the adiabatic deformation. However the $\mathbb{Z}_m$ translation symmetry subgroup, generated by operator $T^n$, is preserved since by construction identical cuts occurs at every $n$th junction. This immediately implies that all the $n$-block states are identical $|\tilde{\Psi}_i\rangle=|\tilde{\Psi}\rangle$ and the total state after the cut is simply $|\tilde{\Psi}\rangle^{\otimes m}$.
Thus we know that the original $\mathbb{Z}_m$ quantum number is the same as the final one which must be trivial since we are dealing with an $n$-block product state $|\tilde{\Psi}\rangle^{\otimes m}$. This implies
\begin{align}
nP(L)=0\mod 2\pi\quad,
\label{eq:LP}
\end{align}
$\forall L=mn$ with $m,n\in\mathbb{Z}^+$ and $n\gg\xi$.

Using this relation on a general system length $L=p_1^{q_1}p_2^{q_2}...p_d^{q_d}$ (here we are using prime factorisation notation) we arrive at the condition
\begin{align}
P(L)=0\mod \frac{2\pi}{p_1^{r_1}p_2^{r_2}...p_d^{r_d}}\quad,
\end{align}
$\forall{r_i\in\{1,...,q_i\}}$ such that $p_1^{r_1}p_2^{r_2}...p_d^{r_d}\gg\xi$. If $L$ factorises into at least two mutually coprime numbers $m,n$ with $m,n\gg\xi$ then these conditions can only be satisfied if 
\begin{align}
P(L)=0\mod 2\pi\quad,
\label{eq:zeroP}
\end{align}
which is satisfied for almost all large enough $L$.

\underline{\textit{Step 2:}} There are a sparse set of cases for which \textit{Step 1} does not enforce trivial momentum, the most notable case being when $\tilde{L}=p^{q}$ with $p$ prime and $q\in\mathbb{Z}^+$. Factorisations such as $\tilde{L}=p_1^{q_1}p_2$ are also not covered if $p_1^{q_1}\not\gg \xi$. 

\begin{figure}[t]
    \centering
        \includegraphics[width=\columnwidth]{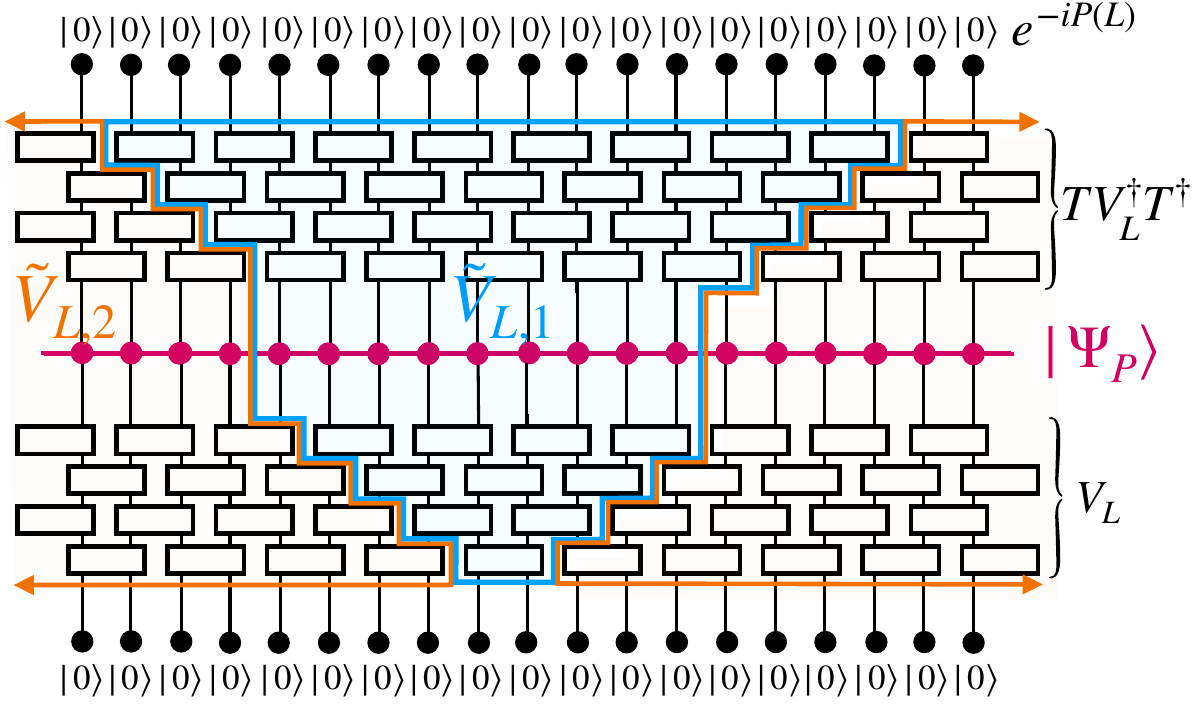}
    \caption{\label{fig:UgeneralL} (Color online) Illustration of splitting $TV_L^\dag T^\dag V_L=\tilde{V}_{L,1}\tilde{V}_{L,2}$ with $\tilde{V}_{L,1}\tilde{V}_{L,2}|\mathbf{0}\rangle=e^{-iP(L)}|\mathbf{0}\rangle$. Here we have taken a snapshot of the circuit to focus on $\tilde{V}_{L,1}$ (framed in blue), however the support of $\tilde{V}_{L,1}$ (in the depicted example 16 qudits) is actually much smaller than the system length. Recall that the circuit is periodic such that the orange arrows, corresponding to components of $\tilde{V}_{L,2}$ (framed in orange), eventually connect on the far side of the ring.
    }
\end{figure}

To show that these cases also possess trivial momentum, once again take a SRE state $|\Psi_{P(L)}\rangle$ on a general length $L$ system with momentum $P(L)\mod 2\pi$. By the definition of a SRE state, there exists a FD quantum circuit $V_{L}$ such that $|\Psi_{P(L)}\rangle=V_L|\mathbf{0}\rangle$. This circuit obeys $TV_LT^\dag|\mathbf{0}\rangle=e^{iP(L)}V_L|\mathbf{0}\rangle$, meaning that it boosts the trivial momentum of the $|\mathbf{0}\rangle$ state by $P(L)\mod 2\pi$. Consider the composition of a circuit
\begin{align}
    \left(TV_L^\dag T^\dag\right) V_L|\mathbf{0}\rangle=e^{-iP(L)}|\mathbf{0}\rangle\quad.
    \label{eq:boostL}
\end{align}
As may be understood via the causality structure the phase $e^{-iP(L)}$ will come piecewise from lightcone circuits. Let us understand this in detail: split $\tilde{V}_L\equiv TV_L^\dag T^\dag V_L$ into a light-cone circuit $\tilde{V}_{L,1}$ and reverse lightcone circuit $\tilde{V}_{L,2}$ such that $\tilde{V}_L=\tilde{V}_{L,1}\tilde{V}_{L,2}$, as depicted in Fig.~\ref{fig:UgeneralL}. The causal structure of the light cone guarantees that a gate $U_1$ in $\tilde{V}_{L,1}$ and a gate $U_2$ in $\tilde{V}_{L,1}$ must commute if $U_2$ appears in a layer after $U_1$, which then allows for the decomposition $\tilde{V}_L=\tilde{V}_{L,1}\tilde{V}_{L,2}$.
Although the exact form of this decomposition is quite malleable, for concreteness let us define $\tilde{V}_{L,1}$ to be constructed causally such that the 1st (lowest) layer consists of a single 2-qudit gate (as seen in Fig.~\ref{fig:UgeneralL}).
$\tilde{V}_{L,1}$ will have support over qudits in the range $[L-\eta,L]$, where by the SRE nature $\eta\ll L$. Due to Eq.~\ref{eq:boostL} we see that
\begin{align}
    \tilde{V}_{L,2} |\mathbf{0}\rangle=|0...0\rangle^{[1,L-\eta-1]}\otimes|\alpha\rangle^{[L-\eta,L]}
\end{align}
for some $|\alpha\rangle$. By construction, we have
\begin{align}
    \tilde{V}_{L,1} |\alpha\rangle=e^{-iP(L)}|0...0\rangle^{[L-\eta,L]}\quad,
    \label{eq:circ1tildeU}
\end{align}
such that we satisfy Eq.~\ref{eq:boostL}.

Now we will extend the circuit $V_L$ from length $L$ to $nL$ for some $n\in\mathbb{Z}^+$, where $n,L$ are coprime and $\gg\xi$, and denote this extended circuit $V_{nL}$. To do this we simply unstitch the circuit $V_L$ at some link and reconnect the ends of $n$ consecutive copies of this unstitched $V_L$ circuit to create a FD quantum circuit $V_{nL}$. Let us see what happens to $\tilde{V}_{nL}\equiv TV_{nL}^\dag T^\dag V_{nL}$ by once again splitting the circuit into two $\tilde{V}_{nL}=\tilde{V}_{nL,1}\tilde{V}_{nL,2}$, where $\tilde{V}_{nL,k}=\prod_{j=0}^{n-1} T^{jL}\tilde{V}_{L,k}(T^\dag)^{jL}$ with $k\in\{1,2\}$. By construction and due to the SRE nature of state construction
\begin{align}
    \tilde{V}_{nL,2} |\mathbf{0}\rangle^{\otimes n}=\left(|0...0\rangle^{[1,L-\eta-1]}\otimes|\alpha\rangle^{[L-\eta,L]}\right)^{\otimes n}\quad.
\end{align}
However, by Eq.~\ref{eq:circ1tildeU}, we have
\begin{align}
    \tilde{V}_{nL,1} \tilde{V}_{nL,2} |\mathbf{0}\rangle^{\otimes n}=e^{-inP(L)}|\mathbf{0}\rangle^{\otimes n}\quad,
    \label{eq:circ1tildeUnL}
\end{align}
so this implies
\begin{align}
    TV_{nL} T^\dag|\mathbf{0}\rangle^{\otimes n} =e^{inP(L)}V_{nL}|\mathbf{0}\rangle^{\otimes n}\quad,
    \label{eq:boostnL}
\end{align}
which means that $V_{nL}$ boosts the momentum of $|\mathbf{0}\rangle$ on a length $nL$ system to a state with momentum $P(nL)=nP(L)\mod 2\pi$. In \textit{Step 1} we showed that $P(nL)=0\mod 2\pi$, so this implies $nP(L)=0\mod2\pi$. Since this holds for two mutually coprime values of $n$, one concludes that $1$d SRE translation-symmetric states have $P(L)=0\mod2\pi$ for all $L\gg\xi$.

\subsection{Higher-dimensional extension}

Our result can be extended to higher dimensions. Consider a $d$-dimensional lattice system and a state $|\Psi\rangle$ that has nontrivial momentum $P$ along, say, $\hat{x}$ direction. We can view the state as a 1d state along the $\hat{x}$ axis, with an enlarged Hilbert space per unit cell (generally exponentially large in $\prod_i L_i$ with $i$ denoting the transverse directions). A finite-depth quantum circuit of the $d$-dimensional system will also be a finite-depth quantum circuit when viewed as a $1$d circuit along the $\hat{x}$-direction (a proof and a somewhat subtle example are presented in Appendix~\ref{app:higherdSRE}; the converse is not true but that does not concern us here). This immediately implies that a SRE state on the $d$-dimensional system must also be SRE when viewed as a $1$d state along $\hat{x}$. What we proved in Sec.~\ref{sec:1dproof} thus implies that the non-trivial momentum state $|\Psi\rangle$ must be long-range entangled. In particular, imposing locality in the transverse directions will only further restrict possible FD circuit, and will certainly not lead to possibilities beyond the $1$d proof. This completes the proof of Theorem~\ref{Thm}.\hfill$\blacksquare$

\subsection{Fermion systems}
\label{sec:Fermions}

It is not difficult to generalize our Theorem~\ref{Thm} to fermionic system. The only subtlety is that the usual definition of translation symmetry in fermion systems has an extra $\mathbb{Z}_2$ sign structure compared to the naive implementation in Eq.~\ref{eq:Tboson}. Instead of specifying the sign structure in the tensor product basis as in Eq.~\ref{eq:Tboson}, it is more convenient to define translation operator through $Tc_{i,\alpha}T^{-1}=c_{i+1,\alpha}$ where $c_{i,\alpha}$ is a fermion operator in unit cell $i$ with some internal index $\alpha$, and $c_{L+1,\alpha}=c_{1,\alpha}$. This operator relation, together with $T|\mathbf{0}\rangle=|\mathbf{0}\rangle$ for the fermion vacuum, uniquely determines the action of $T$ on any state. Now consider a product state $|\varphi\rangle^{\otimes L}$, it is easy to verify that the momentum is $e^{iP}=1$ for odd $L$ and $e^{iP}=\pm1$ for even $L$, where the sign is the fermion parity on each site $\langle\varphi|(-1)^{\sum_\alpha c^{\dagger}_{\alpha}c_{\alpha}}|\varphi\rangle$. We can then go through the proof in Sec.~\ref{sec:proof}, but now with fermion parity preserving FD quantum circuits, and conclude the following:
\begin{theorem}
Any short-range entangled translation eigenstate $|\Psi\rangle$ in a lattice fermion system must have momentum (say in the $x$-direction) $e^{iP_x}=1$ if $L_x$ is odd, and $e^{iP_x}=\pm1$ if $L_x$ is even. States violating this condition must in turn be long-range entangled.
\label{FermionThm}
\end{theorem}
The details of the proof are presented in Appendix~\ref{app:fermionproof}.

Using the same proof technique, we can extend the above result further in various directions. We mention two such extensions without going into the details: (1) for $L_x$ even, the option of $e^{iP_x}=-1$ is possible for a SRE state only if $V/L_x$ is odd ($V=L_xL_y...$ being the volume); (2) if the total fermion parity is odd in a system with even $V$, then any translation eigenstate must be LRE.

\section{Consequences}
\label{sec:consequences}

One of the beauties of Theorem~\ref{Thm} lies in the non-trivial consequences that easily follow. For this section, it is useful to introduce an alternative, but equivalent, formulation of Theorem~\ref{Thm}
\begin{manualtheorem}{1}[Equivalent]
If there exists a finite-depth local unitary that boosts a state's momentum to a different value (mod $2\pi$), then the state is necessarily long-range entangled.
\end{manualtheorem}
The equivalence of this new formulation with the one introduced in Sec.~\ref{sec:intro} can be understood as follows: if all translation-symmetric SRE states possess trivial momentum then non-trivial momentum states must be LRE. Thus if there exists a finite-depth local unitary that can boost a state's momentum to a different value then at least one of either the original or final state possesses non-trivial momentum and must be LRE. The other state is connected to the LRE state via a finite-depth local unitary and thus must also be LRE. The converse follows by contradiction: assume there exists a SRE state that has non-trivial momentum. Such a state (by definition of SRE) is connected via a FD local unitary to the translation-symmetric direct-product state $|\alpha\rangle^{\otimes L}$ which in turn has trivial momentum. Since there now exists a FD local unitary that boosts the momentum to a different value, this implies that the original state was LRE which leads to the contradiction.

This equivalent formulation allows for a direct test for long-range entanglement that we will demonstrate on known and previously unknown LSM theories, and topological orders. In the following discussions we will mostly focus on spin (boson) systems for simplicity, but the results can be generalized quite readily to fermion systems as well.

\subsection{LSMOH constraints}
\label{sec: LSMOH}

The original Lieb-Schultz-Mattis (LSM) theory~\cite{LSM} along with the extensions by Oshikawa~\cite{Oshikawa1999} and Hastings~\cite{Hastings04}, collectively referred to as LSMOH, and their descendants are powerful tools for understanding the low-energy nature of lattice systems. In one of its most potent forms the theorem states that systems with $U(1)$ and translational symmetry that have non-commensurate $U(1)$ charge filling must be `exotic', meaning that they cannot be SRE states. Since the conception of the original LSM theory the field has flourished rapidly with many extensions that impose similar simple constraints based on symmetry~\cite{Nachtergaele2007,Watanabe14551,Po2017,LU2020168060,lu2017liebschultzmattis,PhysRevB.98.125120,PhysRevB.99.075143}, and connections to various fields of physics such as symmetry-protected topological (SPT) order and t'Hooft anomaly in quantum field theory~\cite{Cheng2015,Jian2017,Cho2017,Metlitski2017,PhysRevB.101.224437,Ye2021}. These sort of constraints also have immediate experimental consequences, as they provide general constraints in determining candidate materials of exotic states such as quantum spin liquids~\cite{Balents2010}. Thus, unsurprisingly, there has been a lot of interest in generating more LSMOH-like theorems that provide simple rules to find exotic states. In the following section we provide simple proofs of some known and previously unknown LSMOH theorems.

The first example we consider is the aforementioned non-commensurate 1d $U(1)\times T$ LSM ($T$ denotes the translation symmetry). In this case there exists a local unitary momentum boost that is the large gauge transformation $U=e^{i\frac{2\pi }{L}\sum x\hat{n}_x}$, where $\hat{n}_x$ is the local number operator at $x$. Notice that this transformation is an on-site phase transformation and thus a FD quantum circuit of depth 1. The commutation relation with translation is $TUT^\dag=e^{i 2\pi \frac{\hat{N}}{L}}U$ ($\hat{N}$ being the total charge) which means that for non-commensurate filling $\frac{\hat{N}}{L}\notin\mathbb{Z}$, we may always boost the momentum by a non-trivial value $2\pi \frac{\hat{N}}{L}$. Via the equivalent formulation of Theorem~\ref{Thm}, this immediately implies that non-commensurate filling leads to a LRE state. This observation may be summarised as
\begin{corollary}
{\normalfont($U(1)\times T$ LSM) A $1$d translation and $U(1)$ symmetric state that possesses non-commensurate $U(1)$ charge filling must be long-range entangled.}
\end{corollary}
The standard LSM theorem follows from this statement since we may now apply it to a \textit{ground} state of a $1$d translation and $U(1)$ symmetric local spin Hamiltonian to show that the state must be either gapless or a spontaneously symmetry-broken cat state.
Notice that, strictly speaking, the statement we proved differs slightly from the standard LSM theorem, in that we did not directly prove the vanishing of the energy gap. Rather we showed that any simultaneous eigenstate of translation and $\hat{N}$ such that $\langle\hat{N}\rangle/L\notin\mathbb{Z}$ must be LRE. In principle we do not even need to assume the parent Hamiltonian to be translation or $U(1)$ symmetric, just that the state itself be translation and $U(1)$ symmetric. In fact the statement encompasses all states, not just the ground state, which is perhaps unsurprising since LRE is fundamentally a property of a state and not the Hamiltonian.

The higher-dimensional $U(1)\times T$ LSMOH theorem may be proved following the same logic if $\langle\hat{N}\rangle/L_i\notin\mathbb{Z}$ for some direction $i$ (similar to what was done in Ref.~\cite{Oshikawa1999}). For generic values of $L_i$ the above condition may not hold, and more elaborate arguments are needed (for example see Ref.~\cite{YaoOshikawa2020}) which we will not discuss here. 

Our proof of the LSM theorem has an appealing feature compared to the classic proof~\cite{LSM}: we did not need to show that the state $|\Omega'\rangle=U|\Omega\rangle$ had excitation energy $\sim O(1/L)$ (relative to the ground state $|\Omega\rangle$); rather it suffices for us to show that $|\Omega'\rangle$ has a different lattice momentum compared to $|\Omega\rangle$, which is enough to establish the LRE nature of $|\Omega\rangle$. Next we shall use this simplifying feature to generalize the $U(1)\times T$ LSM theorem to a new constraint involving only discrete $\mathbb{Z}_n\times T$ symmetries.

Let us consider a spin chain ($1$d) with translation symmetry and an onsite $\mathbb{Z}_n$ symmetry generated by $Z\equiv\otimes_iZ_i$ ($Z_i^n=1$). We consider the case when the system size $L=nM$ for some $M\in\mathbb{N}$, and study simultaneous eigenstates of the translation and $\mathbb{Z}_n$ symmetries. If such a state is an unentangled product state $\otimes_i|\varphi\rangle_i$, then by definition $Z=1$ when acting on this state, namely the state carries trivial $\mathbb{Z}_n$ charge. This turns out to be true for any symmetric SRE state, which we now prove. Suppose a translation eigenstate $|\Psi\rangle$ has $Z|\Psi_P\rangle=e^{i2\pi Q/n}|\Psi\rangle$ for some $Q\neq0$ (mod $n$). We can construct a local unitary which is an $\mathbb{Z}_n$-analogue of the large gauge transform
\beq
U=\otimes_iZ_i^i,
\eeq
where $i$ is the unit cell index. For system size $L=nM$ one can verify that $TUT^{-1}U^{\dagger}=Z^{\dagger}$. This means that the momentum of the twisted state $U|\Psi\rangle$ will differ from that of the untwisted $|\Psi\rangle$ by $\langle\Psi|Z^{\dagger}|\Psi\rangle=e^{-i2\pi Q/n}\neq1$. By the equivalent form of Theorem~\ref{Thm} $|\Psi\rangle$ must be LRE. We therefore have
\begin{corollary}
\label{ZnLSM}
{\normalfont($\mathbb{Z}_n\times T$ LSM)} A $1$d translation and $\mathbb{Z}_n$ symmetric ground state that possesses non-trivial $\mathbb{Z}_n$ charge on system lengths $L=nM$ for some $M\in\mathbb{N}$ cannot be short-range entangled, and thus is either gapless or spontaneously symmetry-broken cat state.
\end{corollary}

The above statement also generalizes to higher dimensions if $L_i=nM$ for some direction $i$. For systems with $U(1)$ symmetry, we can choose to consider a $Z_L$ subgroup of the $U(1)$, and the above $\mathbb{Z}_n\times T$ LSM theorem leads to the familiar $U(1)\times T$ LSMOH theorem.

The two LSM-type theorems discussed so far, together with our Theorem~\ref{Thm}, can all be viewed as ``filling-type" LSM theorems, in the sense that these theorems constraint a symmetric many-body state $|\Psi\rangle$ to be LRE when $|\Psi\rangle$ carries certain non-trivial quantum numbers, such as lattice momentum $e^{iP}\neq1$, total $U(1)$ charge $Q\notin L\mathbb{Z}$ or total $\mathbb{Z}_n$ charge $Q\notin L\mathbb{Z}/n\mathbb{Z}$.

There is another type of LSM theorems that involve projective symmetry representations in the onsite Hilbert space, the most familiar example being the spin-$1/2$ chain with $SO(3)$ symmetry. Our Theorem~\ref{Thm} can also be used to understand some (but possibly not all) of the projective symmetry types of LSM. Here we discuss one illuminating example with onsite $\mathbb{Z}_2\times\mathbb{Z}_2$ symmetry in one dimension~\cite{PhysRevB.83.035107,Ogata2019,Ogata2021}, such that the generators of the two $\mathbb{Z}_2$ group anti-commutes when acting on the local Hilbert space: $X_iZ_i=-Z_iX_i$ (this can simply be represented by the Pauli matrices $\sigma_x$, $\sigma_z$). Now set the length $L=2N$ with odd $N$, and consider the three local unitaries $U_x=(\mathds{1}\otimes\sigma_x)^{\otimes N}$, $U_z=(\mathds{1}\otimes\sigma_z)^{\otimes N}$, and $U_{xz}=(\sigma_x\otimes\sigma_z)^{\otimes N}$. One can verify the commutation relations $TU_{x}T^\dag=(-1)^{Q_x}U_x$, $TU_{z}T^\dag=(-1)^{Q_z}U_z$, and $TU_{xz}T^\dag=(-1)^{N+Q_x+Q_z} U_{xz}$. These commutation relations imply that the momentum of any symmetric state $|\Psi\rangle$ will be boosted by $\Delta P=\pi$ by at least one of the three unitaries, therefore $|\Psi\rangle$ must be LRE by Theorem~\ref{Thm}.

\subsection{Topological orders: weak CDW}
\label{sec:wCDW}

We now consider an intrinsic (bosonic) topological order on a $d$-dimensional torus. By definition there will be multiple degenerate ground states, separated from the excitation continuum by a finite energy gap. If one of the ground states $|\Psi_a\rangle$ has a non-trivial momentum, say along the $\hat{x}$ direction, then according to Theorem 1 this state should be LRE even when viewed as a one-dimensional system in $\hat{x}$ direction (with the other dimensions $y,z...$ viewed as internal indices). Since there is no intrinsic topological order in one dimension, the only mechanism for the LRE ground state is spontaneous symmetry breaking. The lattice translation symmetry is the only relevant symmetry here -- all the other symmetries can be explicitly broken without affecting the LRE nature of $|\Psi_a\rangle$, since the state will still have nontrivial momentum. Therefore $|\Psi_a\rangle$ must be a cat state that spontaneously breaks the $\hat{x}$-translation symmetry~\footnote{Another way to see this is to note that a cat state is composed of individual SRE states. Since we have proven that translation symmetric SRE states possess trivial momentum, it follows that the cat state may only achieve non-trivial momentum when the individual SRE states break translation symmetry, i.e. the cat state must correspond to translation symmetry breaking.}, also known as a charge density wave (CDW) state~\cite{RevModPhys.60.1129}. Furthermore, any other ground state $|\Psi_{b\neq a}\rangle$ can be obtained from $|\Psi_a\rangle$ by a unitary operator $U_{ba}$ that is non-local in the directions transverse to $\hat{x}$, but crucially is local in $\hat{x}$ -- for example in two dimensions $U_{ba}$ corresponds to moving an anyon around the transverse cycle. By Theorem~\ref{Thm} we then conclude that $|\Psi_b\rangle$ is also a CDW in $\hat{x}$. 

Perhaps the most familiar example of the above statement is the fractional quantum Hall effect. It is known that the $1/k$ Laughlin state on the torus is adiabatically connected to a quasi-one-dimensional CDW state in the Landau gauge, also known as the Tao-Thouless state~\cite{PhysRevB.28.1142,PhysRevB.77.155308}. For example for $k=2$ the Tao-Thouless state with momentum $P=\pi n$, in the Landau orbit occupation number basis, reads
\begin{equation}
    |101010...\rangle+e^{i\pi n}|010101...\rangle.
\end{equation}

The CDW nature of the ground states is perfectly compatible with the topological order being a symmetric state, since there is no \textit{local} CDW order parameters with nonzero expectation value. The CDW order parameter in this case is non-local in the directions transverse to $\hat{x}$. For example, in two-dimensions the CDW order parameter is defined on a large loop that wraps around the cycle transverse to $\hat{x}$. This phenomenon is dubbed \textit{weak} symmetry breaking in Ref.~\cite{Kitaev06}. The weak spontaneous symmetry breaking requires a certain degeneracy for the ground state. This degeneracy is naturally accommodated by the ground state manifold of the topological order. For example for the above Tao-Thouless state at $k=2$ the CDW order requires a two-fold ground state degeneracy, which is nothing but the two degenerate Laughlin states on torus.

The above results can be summarized as follows:
\begin{corollary}
\label{cor:toporderCDW}
If a ground state of a gapped topological order on a $d$-dimensional torus ($d>1$) has a non-trivial momentum in $\hat{x}$, then any ground state of this topological order must \textit{weakly} break translation symmetry in $\hat{x}$.
\end{corollary}
A further example of these results, alongside the effects of anyon condensation, applied upon the $\mathbb{Z}_2$ topologically ordered Toric code is demonstrated in the Appendix~\ref{app:Toriccode}. The above result also implies the following constraint on possible momentum carried by a topologically ordered ground state:
\begin{corollary}
If a gapped topological order has $q$ degenerate ground states on torus, then the momentum of any ground state in any direction is quantized: 
\begin{equation}
P^{(a)}_i=2\pi N^{(a)}_i/q,
\end{equation}
where $N^{(a)}_i$ is an integer depending on the ground state (labeled by $a$) and direction $i$.
\end{corollary}
This is simply because for other values of the momentum, the ground state degeneracy required by the spontaneous translation-symmetry-breaking order will be larger than the ground state degeneracy from the topological order, which results in an inconsistency. An immediate consequence of the above corollary is that invertible topological orders (higher-dimensional states that are LRE by our definition but has only a unique gapped ground state on closed manifolds), such as the chiral $E_8$ state\cite{Kitaev06}, cannot have nontrivial momentum on a closed manifolds since $q=1$.

The above statement immediately implies that the momenta of topological ordered ground states are robust under adiabatic deformations, as long as the gap remains open and translation symmetries remain unbroken. For the Tao-Thouless states this conclusion can also be drawn from the LSM theorem if the $U(1)$ symmetry is unbroken. Our result implies that the momenta of Laughlin-Tao-Thouless states are robustly quantized even if the $U(1)$ symmetry is explicitly broken.

\subsection{Crystalline symmetry-protected topological phases}
\label{sec:cSPT}

There has been growing interest and successes in understanding the symmetry-protected topological (SPT) phases associated with crystalline symmetries~\cite{PhysRevX.7.011020,ThorngrenElse,PhysRevB.96.205106,Shiozaki2018,SongFangQi2018,Else2018}. When the protecting symmetry involves lattice translation, a crucial ``smoothness'' assumption~\cite{ThorngrenElse,PhysRevB.96.205106} is used. Essentially one assumes that for such SPT phases the inter unit-cell entanglement can be adiabatically removed, possibly with the help of additional ancilla degrees of freedom. This allows one to formally ``gauge'' the translation symmetry~\cite{ThorngrenElse} and build crystalline topological phases out of lower-dimensional states~\cite{PhysRevB.96.205106,SongFangQi2018,Else2018}. 

Our result, namely Theorem~\ref{Thm}, serves as a non-trivial check on the smoothness assumption in the following sense. If there were SRE states with non-trivial lattice momenta, such states would have irremovable inter-unit cell entanglement since unentangled states cannot have non-trivial momentum. Equivalently the correlation length $\xi$ cannot be tuned to be smaller than the unit cell size $a$. In fact, if such states exist, they would by definition be non-trivial SPT states protected solely by translation symmetry -- such SPT states would be beyond all the recent classifications. 

We note that our result is a necessary condition, but not a proof, for the smoothness assumption, as there may be other ways to violate the assumption without involving a ground state momentum. It will be interesting to see if the arguments used in this work can be extended to fully justify the smoothness assumption.

\section{Discussions}
\label{sec:discussions}

In this paper we have shown that a quantum many-body state with non-trivial lattice momentum is necessarily long-range entangled, hence establishing a simple yet intriguing connection between two extremely familiar concepts in physics: translation symmetry and quantum entanglement. Many directions can be further explored, which we briefly comment on in the remainder of this Section.

One important aspect that we have so far skipped over is that LSM theory is in fact intimately connected to quantum anomalies~\cite{Cheng2015,Jian2017,Cho2017,Metlitski2017,PhysRevB.101.224437,Ye2021}. This is natural since they both provide UV conditions that constraint the low-energy behaviours. For the ``projective symmetry" type of LSM theorems, this connection has been precisely established and it is known that such LSM constraints correspond to certain discrete (quantized) t'Hooft anomalies. For the ``partial filling'' type of LSM such as the familiar $U(1)\times T$ constraint, however, the connection has been discussed~\cite{Song2021Polarization,Else2021FL,Wang20,PhysRevResearch.3.043067} but has yet to be fully developed. As we discussed in Sec.~\ref{sec: LSMOH}, our main result (Theorem~\ref{Thm}) can be viewed as a ``partial filling'' type of LSM that only involves translation symmetry. It is therefore natural to ask whether Theorem~\ref{Thm} can be understood from an anomaly perspective. To achieve this goal, it is clear that the standard quantized t'Hooft anomaly is insufficient (a point which was also emphasized in Ref.~\cite{Else2021FL} for the $U(1)\times T$ LSM) -- for example, the toric code discussed in Appendix~\ref{app:Toriccode} has no t'Hooft anomaly since one can condense the $e$ particle to obtain a trivial symmetric state. One would therefore need to expand the notion of anomaly to accommodate the partial-filling type of LSM constraints including the one discussed in this work, possibly along the line of the ``unquantized anomaly" discussed in Ref.~\cite{PhysRevResearch.3.043067}. We leave this aspect to future work.

Another powerful consequence of the traditional $U(1)\times T$ LSM theorem is on the stability of the LRE ground states (with partial charge filling) under symmetric perturbations: assuming the charge compressibility is finite (could be zero), then a small perturbation will not change the charge filling discontinuously, so the system remains LRE under small symmetric perturbations (unless the perturbation leads to spontaneous symmetry-breaking like the BCS attraction). It is natural to ask whether the other ``partial filling" types of LSM theorems can serve similar purposes. In fact Ref.~\cite{PhysRevResearch.3.043067} discussed precisely this point under the notion of ``unquantized anomaly''. The unquantized anomalies are very similar to Theorem~\ref{Thm} and \ref{FermionThm} and Corollary~\ref{ZnLSM}, except that the key quantity is not the discrete charges (lattice momentum or $\mathbb{Z}_n$ charges) on a specific systems size $L$, but the charge densities (momentum density or $\mathbb{Z}_n$ charge density). Such discrete charge densities can not be defined for a fixed $L$, but may be defined for a sequence of systems with $L\to \infty$. Ref.~\cite{PhysRevResearch.3.043067} argued that, in the context of Weyl and Dirac semimetals, as long as these discrete charge densities are well behaved in the $L\to\infty$ limit, the unquantized anomalies will protect the LRE nature of the states under symmetric perturbations. Our work here can be viewed as a rigorous justification of the unquantized anomalies in Ref.~\cite{PhysRevResearch.3.043067} on fixed system sizes.

Assuming a well-behaving momentum density in the thermodynamic limit, we can also apply our results to a Fermi liquid with a generic Fermi surface shape, such that the ground state from the filled Fermi sea has a non-vanishing momentum density (this requires breaking of time-reversal, inversion and reflection symmetries). This can be viewed as a non-perturbative explanation for the stability of such low-symmetry Fermi surface, even in the absence of the charge $U(1)$ symmetry. (Recall that perturbatively the stability comes from the fact that the Cooper pairing terms no longer connect opposite points on the Fermi surface).

Another question one may ask is whether a broader group of non-onsite symmetries obey similar charge and entanglement restrictions. It is easy to see that exactly the same constraint holds for glide reflections and screw rotations, since when the system is viewed as $1$d there is no difference between glide reflection, screw rotation and translation. It is also easy to see that the constraint does \textit{not} hold for point group symmetries (rotations and reflections), because such symmetries will be onsite at some points in space (the fixed points of point groups). It is therefore important that translation symmetry is \textit{everywhere} non-onsite. The question becomes even more intriguing if we consider more general unitary operators (such as quantum cellular automata~\cite{GNVW}).

There are many more natural avenues for further exploration. The interplay between the non-local nature of translation symmetry with crystalline symmetry anomalies is not yet well understood and requires more concrete mathematical grounding such as a rigorous proof of when the smoothness condition is valid. Relatedly it remains to be determined whether translation symmetry may be truly treated as an onsite symmetry and gauged, or whether its non-locality and non-trivial momentum may hinder or require modifications to the usual gauging process. Implications of our results on the ``emergibility" of phases may also provide fruitful insights to achievable and unachievable states on the lattice~\cite{PhysRevX.11.031043,Ye2021}. Our work has shown without a doubt that translation symmetry is many-faceted and plays a crucial role in the entanglement properties of crystalline materials.

\begin{acknowledgments}
We acknowledge insightful discussions with Yin-Chen He, Timothy Hsieh, and especially Liujun Zou. We thank Anton Burkov for a previous collaboration that inspired this work. We thank the anonymous referees for their careful reading of our manuscript and their illuminating comments and questions. LG was supported by the Natural Sciences and Engineering Research Council (NSERC) of Canada and by a Vanier Canada Graduate Scholarship. 
Research at Perimeter Institute is supported in part by the Government of Canada through the Department of Innovation, Science and Economic Development and by the Province of Ontario through the Ministry of Economic Development, Job Creation and Trade.
\end{acknowledgments}

\appendix

\section{Higher dimensional FD quantum circuit}
\label{app:higherdSRE}

Here we will show that a higher dimensional ($\mathrm{d}>1$) FD quantum circuit viewed in 1d (say along $\hat{x}$), where each enlarged unit cell Hilbert space is now exponential in the transverse dimension $\prod_i L_i$, is also a FD quantum circuit.

\begin{figure}[ht]
    \centering
        \includegraphics[width=\columnwidth]{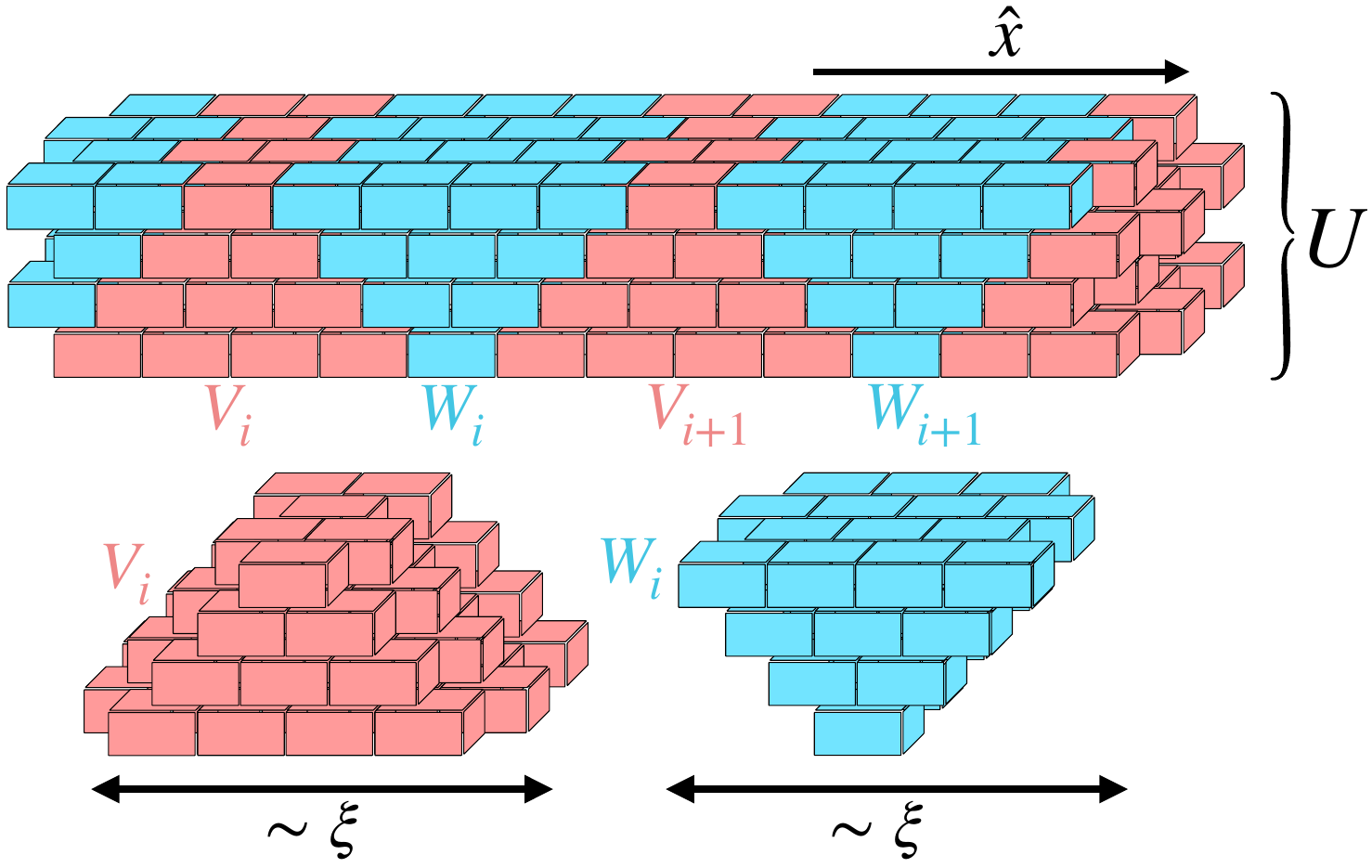}
    \caption{\label{fig:higherFDcircuit} (Color online) A sample 2d FD quantum circuit $U$ decomposed along $\hat{x}$ into `extended lightcone' unitaries $\{V_{i}\}$ (shaded red) and `extended reverse lightcone' unitaries $\{W_{i}\}$ (shaded blue). The exact position to begin the lightcone cut is variable, although here we have done so symmetrically.}
\end{figure}

To see this, let us decompose the higher dimensional FD quantum circuit $U$ into two sets of unitaries via `zig-zag' cuts following lightcone pathways along $\hat{x}$. We depict an example of such a cut applied to a 2d FD quantum circuit $U$ in Fig.~\ref{fig:higherFDcircuit}. The two sets of unitaries consist of self-commuting `extended lightcone' unitaries $\{V_{i}\}$ and a set of self-commuting `extended reverse lightcone' unitaries $\{W_{i}\}$, such that $U=\prod_i W_i\prod_j V_j$. Due to the finite correlation length $\xi$ in SRE systems and correlations necessarily arise from the lightcone structure, each unitary component spans $\sim \xi\ll L$ unit cells in $\hat{x}$. This decomposition forms a 1d FD quantum circuit with two layers ($\{V_{i}\}$ and $\{W_{i}\}$).

Thus we have shown that higher-dimensional SRE states remain SRE when viewed in 1d.

We now discuss a somewhat subtle example to further illustrate the point\footnote{We thank an anonymous referee for raising this example.}. Consider a $(2+1)d$ system of fermions with global symmetry $\mathbb{Z}_2\times \mathbb{Z}_2^f$ where $\mathbb{Z}_2^f$ is the fermion parity conservation. Essentially we have two flavors of fermions, one that transforms trivially under the global $\mathbb{Z}_2$ and another that transforms with a minus sign. Now put the $\mathbb{Z}_2$-even fermion in a $p+ip$ superconductor and the $\mathbb{Z}_2$-odd fermion in a $p-ip$ superconductor. It seems natural to consider this state SRE since the state can be trivialized by breaking the $\mathbb{Z}_2$ symmetry (which can be seen, for example, by examining the edge states). However, it turns out that when put on torus, the state is strictly SRE only if the two fermions ($\mathbb{Z}_2$ even and odd) have the same boundary conditions in space. If the two fermions have opposite boundary conditions -- say one with periodic and the other with anti-periodic boundary conditions in $\hat{y}$, then when viewed as a one-dimensional system along $\hat{x}$, the system forms a Kitaev chain with unpaired Majorana zero modes at the ends. Crucially, a Kitaev chain does not require any global symmetry (besides the $\mathbb{Z}_2^f$ which is anyway un-breakable) and therefore cannot be adiabatically connected to a trivial state. Such ``invertible'' topological state is considered LRE in the definition adopted in this work. So by simply twisting the boundary condition in $\hat{y}$ direction, we have converted a SRE state to a LRE one!

The above example in fact does not contradict our result in this section. What it really shows is that the $\mathbb{Z}_2$ boundary condition cannot be twisted adiabatically for this state. Namely, there is no adiabatic path (FD quantum circuit) that can change the boundary condition (along a space cycle) from $\mathbb{Z}_2$-periodic to $\mathbb{Z}_2$-anti-periodic. Indeed the most familiar adiabatic operation that could change the boundary condition (say in $\hat{y}$) involves creating two $\mathbb{Z}_2$ vortices, moving one in $\hat{x}$ across the entires system, and re-annihilate with the other one at the end. This process requires a time scale (or circuit depth) of order $O(L_x)$. One may wonder if a more clever construction can bring the circuit depth down to $O(1)$, but the previous discussion on the SRE vs. LRE nature shows that this is impossible\footnote{If the symmetry is not $\mathbb{Z}_2$ but a continuous one such as $U(1)$, the twist can be achieved adiabatically by slowly threading a (continuous) gauge flux. So for continuous symmetries we do not expect the SRE vs. LRE nature to change under twisted boundary conditions -- this is indeed compatible with the fact that $p\pm ip$ superconductors are not compatible with $U(1)$ symmetry.}.

\section{Proof for fermion systems}
\label{app:fermionproof}

In this section we will carefully go through the $1$d proof for fermionic systems. The key difference between fermionic and bosonic systems, as discussed in Sec.~\ref{sec:Fermions}, is that a product state in fermion system has momentum $P=0$ mod $\pi$ instead of mod $2\pi$ for bosons. More specifically, for fermion systems with odd system length, all product states have zero momentum, just like the bosonic case. However for even system length, product states may either have zero or $\pi$ momentum depending on the fermion parity per site being even or odd respectively.\footnote{Here we demand the translationally-invariant product state to be an eigenstate for the fermion parity in each unit cell. This is because the state must be an eigenstate of the total fermion parity, as required by the fermion parity superselection rule.} Additionally we note that for even system length all translation symmetric product states possess even total fermion parity.

Fermionic local unitaries are defined via fermion parity preserving Hamiltonians~\cite{PhysRevB.91.125149}, and thus can only be represented via parity preserving FD quantum circuits. More specifically, since parity is an on-site symmetry, each unitary that makes up the parity preserving FD quantum circuit must themselves be parity preserving. The proof in Sec.~\ref{sec:1dproof} directly carries over for SRE fermionic systems, but we must now keep in mind the system size and total fermion parity of the system. These initial conditions lead to different possibilities dependent on the achievable translation symmetric fermionic product states, as alluded to above.

The proof for odd length fermionic systems for both even and odd total fermion parity follows step by step with the bosonic proof. For example, in \textit{Step 1} there is no trouble with `cutting' a system of length $L=mn$ into $m$ segments of length $n$ since the amount of segments ($m$) will still be odd if $L$ is odd. The resulting odd number of segments implies trivial momentum when translating by $n$ such that $nP(L)=0\mod 2\pi$, just as in the bosonic case. Similarly, in \textit{Step 2}, we may always glue $n$ amounts of length $L$ segments with $n$ being odd; $nL$ will still be odd so that we may then apply \textit{Step} 1 to arrive at the same conclusion that $P(L)=0\mod 2\pi$.

The story is slightly more complicated for even length fermionic systems. Let us first consider the even total parity case. Here, in \textit{Step 1} we must be careful when we cut the $L=mn$ length system into $m$ length $n$ segments. If $m$ is even, then we have the condition $nP(L)=0\mod \pi$ (note that this is $\pi$ instead of $2\pi$). If $m$ is odd, then $n$ must be even, and we have the condition $nP(L)=0\mod 2\pi$. If $L$ is divisible by two mutually-coprime numbers $p_1\gg\xi$ and $p_2\gg\xi$, i.e. $L=p_1p_2p_3$ for some $p_3\in\mathbb{Z}^+$, then we have two scenarios: 1. One is even, say $p_1$, and one is odd, say $p_2$, such that we have $p_1P(L)=0\mod \pi$, $p_2P(L)=0\mod 2\pi$ for which we conclude $P(L)=0\mod \pi$; 2. Both $p_1,p_2$ are odd (in this case $p_3$ will be even such that $L$ is even), then we have $p_1P(L)=0\mod \pi$ and $p_2P(L)=0\mod \pi$ such that $P(L)=0\mod \pi$. So the best condition we may arrive at is $P(L)=0\mod \pi$, as opposed to $P(L)=0\mod 2\pi$ in the bosonic case. For length that \textit{Step 1} does not cover (e.g. $L=2p$ where $p$ is a prime and $2\ll\xi$), we again turn to \textit{Step 2}. Here the proof for the bosonic case applies with a minor alteration that in the final step, after the glueing procedure, we can only conclude via \textit{Step 1} that $nP(L)=0\mod \pi$. Choosing two mutually-coprime values for $n$, we may then conclude that $P(L)=0\mod\pi$ for all $L$. Here we may intuitively gain an understanding of the $\mathrm{mod}\,\pi$ factor from observing the translation symmetric product states with even total fermion parity on even system lengths: the fermion vacuum state $|\mathbf{0}\rangle$ possesses zero momentum and a state with one fermion per site, say $\prod_{i=1}^Lc^{\dagger}_i|\mathbf{0}\rangle$ possesses $\pi$ momentum. The two states can be related to one another via a fermionic FD quantum circuit, e.g. a layer of $|0\rangle_i\otimes|0\rangle_{i+1}\rightarrow c^{\dagger}_ic^{\dagger}_{i+1}|0\rangle_i\otimes|0\rangle_{i+1}$ operators. This indicates that for SRE states zero and $\pi$ momentum may be adiabatically connected with each other, thus leading to $\mathrm{mod}\,\pi$ rather than $\mathrm{mod}\,2\pi$.

The story is drastically different for even length systems with odd total fermion parity. Here, the cutting procedure in \textit{Step 1} leads to a contradiction: for $L=mn$ with $m,n\gg\xi$, $m$ and/or $n$ must be even. Let $m$ be even such that we may create a FD quantum circuit that can divide the system into $m$ identical segments of length $n$. However since $m$ is even and the segments are identical then the total fermion parity must be even. This contradicts our initial assumption, so we must conclude that the initial state cannot be SRE, i.e. a FD quantum circuit that divides the system cannot be created since the FD quantum circuit $U:|\Psi_{P(L)}\rangle\rightarrow |\mathbf{0}\rangle$ does not exist. For lengths that \textit{Step 1} does not cover, we may again apply the logic of \textit{Step 2} to arrive at a contradiction: if $|\Psi_{P(L)}\rangle$ is SRE then, via the glueing procedure, we may construct a FD quantum circuit for length $nL$ with odd $n$ such that the total fermion parity is still odd. $n$ may always be chosen such that $nL=\tilde{n}\tilde{m}$ with $\tilde{m}$ is even and $\tilde{m}\gg\xi$ so we may again create a circuit that divides the system into $\tilde{m}$ identical segments of length $\tilde{n}$, from which we conclude that the total fermion parity is even. By contradiction, this means that all translation symmetric even length fermionic systems with odd total fermion parity must be LRE. Intuitively this may be understood since there exist no even length translation symmetric product states with odd total fermion parity.

\section{Weak CDW example - Toric code}
\label{app:Toriccode}

In this section we will demonstrate the effects of weak translation symmetry breaking and anyon condensation on a well-known $\mathbb{Z}_2$ topological order example: the Toric code with a gauge charge at each lattice site.

Take such a modified Toric code on a square lattice $L_x\times L_y$ with even $L_x$ and odd $L_y$ and periodic boundary conditions, with a Hamiltonian given by
\begin{align}
    H=J_e\sum_+\prod_{i\in+} \sigma_x - J_m \sum_\square\prod_{i\in\square} \sigma_z\quad,
    \label{eq:ToricCode}
\end{align}
where $J_e,J_m>0$. Here we have chosen a positive sign in front of the $+$ term instead of the usual negative sign, which corresponds to a configuration with a gauge charge (``e'' anyon) at each lattice site. By construction this system respects translation symmetry in $x$ and $y$, enacted by operators $T_x$ and $T_y$.

Let us define large cycle electric and magnetic charge operators $V_x=\prod_{\bar{C}_x}\sigma_x$, $V_y=\prod_{\bar{C}_y}\sigma_x$, $W_x=\prod_{C_x}\sigma_z$, $W_y=\prod_{C_y}\sigma_z$, where $C_{x,y}$ are given by cycles along the lattice links in the $x/y$ directions and $\bar{C}_{x,y}$ are cycles in the $x/y$ direction that are perpendicular to the lattice links. Physically these operators correspond to creating anyon pairs (charge ``e'' excitations for $W$ operators and flux ``m'' excitations for $V$ operators), moving one of the anyons along the relevant cycle and then re-annihilating the anyons.

The degenerate ground states of such a system may be derived from the following translation symmetric topological ground state
\begin{align}
    |\mathbf{0}\rangle=(1+V_x)\prod_+(1-\prod_{i\in +} \sigma_x)|\Uparrow\rangle\quad,
\end{align}
where $|\Uparrow\rangle$ is the all spin-up state and for simplicity we have ignored normalization. This state is an eigenstate of $V_x$ and $W_x$ with eigenvalue $|v_x,w_x\rangle=|1,1\rangle$.
Due to the relations
\begin{align}
    W_xV_y=-V_yW_x\quad, && W_yV_x=-V_xW_y\quad,\\
    T_xV_y|\mathbf{0}\rangle=-V_yT_x|\mathbf{0}\rangle\quad, && 
    T_xW_y|\mathbf{0}\rangle=W_yT_x|\mathbf{0}\rangle\quad,
\end{align}
operators $V_y$ and $W_y$ will allow us to generate the remaining ground states $\{|\mathbf{0}\rangle,V_y|\mathbf{0}\rangle=|1,-1\rangle,W_y|\mathbf{0}\rangle=|-1,1\rangle,V_yW_y|\mathbf{0}\rangle=|-1,-1\rangle\}$ of which $V_y|\mathbf{0}\rangle$ and $V_yW_y|\mathbf{0}\rangle$ have an $\hat{x}$ momentum boost of $\pi$ compared to the other two ground states. Thus, by Corollary~\ref{cor:toporderCDW}, all ground states must weakly break translational symmetry. How can we see this more concretely?

\begin{figure}[ht]
    \centering
        \includegraphics[width=\columnwidth]{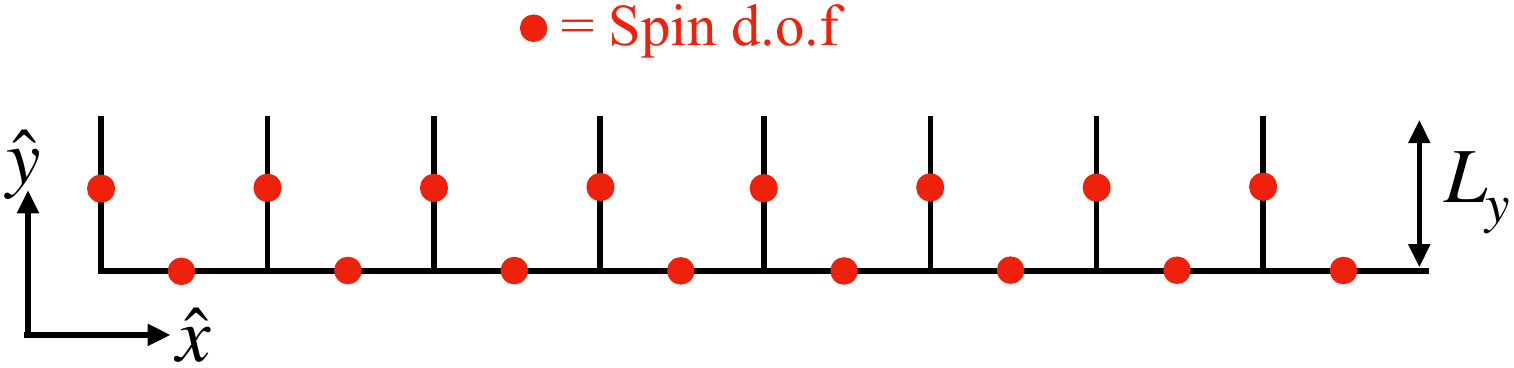}
    \caption{\label{fig:toric1dchain} (Color online) The Toric code system on a periodic lattice with $L_y=1$. There are two spin degrees of freedom (d.o.f) per unit cell in $\hat{x}$.}
\end{figure}

The easiest way to see this CDW effect is to take this Toric code system with $L_y=1$. Since all topological information is contained in a single plaquette, such a system is topologically no different compared to a general odd $L_y$ system. With $L_y=1$, the system reduces to two spins per unit cell in the $\hat{x}$ direction, which we depict in Fig.~\ref{fig:toric1dchain}.

Here the Toric code Hamiltonian reduces to
\begin{align}
    H=J_e\sum_{i\in-} \sigma_x^{[i]} \sigma_x^{[i+1]} - J_m \sum_{ i\in |} \sigma_z^{[i]}\sigma_z^{[i+1]}\quad,
    \label{eq:1dToricCode}
\end{align}
where the $i$ sum is over the horizontal lattice sites (denoted $-$) for the first term and over vertical lattice sites (denotes $|$) for the second term. This Hamiltonian simply describes two decoupled Ising chains, where the first term is in an antiferromagnetic state while the second term is in a ferromagnetic state. The ground states for the respective chains are $\{|\Rightarrow\Leftarrow\rangle\equiv|\rightarrow\leftarrow\rightarrow\leftarrow ...\rightarrow\leftarrow\rangle,|\Leftarrow\Rightarrow\rangle\equiv|\leftarrow\rightarrow\leftarrow\rightarrow ...\leftarrow\rightarrow\rangle\}$ and $\{|\Uparrow\rangle\equiv|\uparrow\uparrow...\uparrow\rangle,|\Downarrow\rangle\equiv|\downarrow\downarrow...\downarrow\rangle\}$. The four ground states of the total system are thus given by $\{|\Rightarrow\Leftarrow\rangle|\Uparrow\rangle,|\Rightarrow\Leftarrow\rangle|\Downarrow\rangle,|\Leftarrow\Rightarrow\rangle|\Uparrow\rangle,|\Leftarrow\Rightarrow\rangle|\Downarrow\rangle\}$. It is clear that these correspond to CDW states since the antiferromagnetic Ising chain breaks the $\hat{x}$ directional translational $\mathbb{Z}_{L_x}$ symmetry group to $\mathbb{Z}_{L_x/2}$. Relating back to our original ground (``cat'') state notation, we have
\begin{align}
    |\mathbf{0}\rangle&=|\Rightarrow\Leftarrow\rangle|\Uparrow\rangle+|\Rightarrow\Leftarrow\rangle|\Downarrow\rangle\nonumber\\&+|\Leftarrow\Rightarrow\rangle|\Uparrow\rangle+|\Leftarrow\Rightarrow\rangle|\Downarrow\rangle\quad,
\end{align}
\begin{align}    
    V_y|\mathbf{0}\rangle&=|\Rightarrow\Leftarrow\rangle|\Uparrow\rangle+|\Rightarrow\Leftarrow\rangle|\Downarrow\rangle\nonumber\\&-|\Leftarrow\Rightarrow\rangle|\Uparrow\rangle-|\Leftarrow\Rightarrow\rangle|\Downarrow\rangle\quad,
\end{align}
\begin{align}
     W_y|\mathbf{0}\rangle&=|\Rightarrow\Leftarrow\rangle|\Uparrow\rangle-|\Rightarrow\Leftarrow\rangle|\Downarrow\rangle\nonumber\\&+|\Leftarrow\Rightarrow\rangle|\Uparrow\rangle-|\Leftarrow\Rightarrow\rangle|\Downarrow\rangle\quad,
\end{align}
\begin{align}
     V_yW_y|\mathbf{0}\rangle&=|\Rightarrow\Leftarrow\rangle|\Uparrow\rangle-|\Rightarrow\Leftarrow\rangle|\Downarrow\rangle\nonumber\\&-|\Leftarrow\Rightarrow\rangle|\Uparrow\rangle+|\Leftarrow\Rightarrow\rangle|\Downarrow\rangle\quad,
\end{align}
so we see that the toric code ground states indeed corresponds to weak translation symmetry breaking with non-local order parameter $\langle V_y\rangle$, which when viewed in 1d along $\hat{x}$ can be interpreted as a \textit{local} order parameter.

Let us now consider the effects of anyon condensation, and the phases that it may lead to. On general ground we expect that a symmetric, confined state could arise from condensing the $e$ anyon, since the $e$ anyon does not carry any nontrivial projective quantum number in the toric code model. This means that when viewed as a $1d$ system, the $e$-condensation gives a transition between the CDW phase and a symmetric phase. Since the CDW phase has two-fold symmetry breaking, it is natural to expect that the transition is simply of the Ising type. These phenomena can be easily demonstrated in the $L_y=1$ limit, as we describe as follows. To drive condensation of the e anyons, we may add a $-h_e\sum_{i}\sigma_z$ term to Eq.~\ref{eq:ToricCode}. For the horizontal bonds this simply leads to a transverse-field Ising model. For $h_e<J_e$, we will still be in the topological ordered state, for $h_e=J_e$ we will have the Ising critical point, and for $h_e>J_e$ we are in the disordered (trivial) state. In the $L_y=1$ example, increasing $h_e$ corresponds to transitioning to the $|\Uparrow\rangle$ phase for the antiferromagnetic Ising chain, which has restored the $\hat{x}$ directional translation symmetry to give the trivial symmetric state.

Similarly we may try to condense the type m anyons by adding a $-h_m\sum_{i}\sigma_x$ term to Eq.~\ref{eq:ToricCode}. In this case the anyon behaves non-trivially under either $T_x$ or $T_y$ since they anticommute when acting on m. Condensing the anyon leads to symmetry breaking of either $T_x$ or $T_y$ (dependent on the specific energetics of the system), so we expect to transition to a true $2$d cat (i.e. symmetry-broken) state. The effects of this condensation cannot be readily seen on the $L_y=1$ lattice example since translation in $\hat{y}$ ceases to have meaning. However it is known that such a translation symmetry-breaking transition occurs such that the final state forms a valence bond solid~\cite{Sachdev_20181}.

\bibliography{references}

\end{document}